\documentstyle[aps,psfig]{revtex} 
\begin{document} 
\title{Implementation of parallel algorithms for 2D vortex dynamics simulation in type-II 
superconductors} 
\author{Mahesh Chandran} 
\address{ Department of Condensed Matter Physics and Material Sciences, \\ Tata Institute
of Fundamental Research, Mumbai 400 005.\\} 
\date{22 December 2000} 
\maketitle 
\nopagebreak
\pagenumbering{arabic} 
\begin{abstract} 
This report discusses the implementation of two parallel
algorithms on a distributed memory system for studying vortex dynamics in type-II superconductors.
These algorithms have been previously implemented for classical molecular dynamics simulation with
short-range forces\cite{plimp}. The run time for parallel algorithm is tested on a system
containing upto 4 processors and compared with that for vectorized algorithm on a single processor
for system size ranging from 120 to 4800 vortices.  
\end{abstract}

\section{Introduction}

The classical molecular dynamic (MD) simulation is a widely used computational method for studying
the collective behaviour of interacting particles and appears in many diverse areas of science. The
MD simulation involves numerically solving the classical Newton's equation of motion for each
particle starting from a set of initial conditions. The force on each particle in derived from 
the potential energy functional which contains the physics of the problem. Since the equation 
of motion of individual particle is solved, the MD simulation allows one to obtain details of 
the static and dynamical properties at the smallest length scale in the problem.

The computational complexity of the MD lies in two factors : the number of particles $N$, and the
time scale of the phenomenon under simulation. A typical example is the melting of a simple cubic
solid with dimension 50 lattice constants which requires 125,000 atoms simulation. Also, 
simulation of few pico-seconds of the real time dynamics would correspond to thousands of 
time steps which can be prohibitive even for large computers. Considerable research has therefore gone into 
optimizing algorithms for MD simulation depending upon the problem in hand. With development of 
parallel machines, various scalable algorithms for MD simulation has been given which can be used 
on a system with few processors to hundreds of processors. These algorithms use the message-passing
model of programming for inter-processor communication. Another advantage of writing simulation
code using message passing is due to availability of standard library which allows portability to
various computing platforms.

In this report, we describe the implementation of two parallel MD algorithms for simulating 2D
vortex dynamics in type-II superconductors. The algorithm used here have been previously developed
and implemented by others for simulating short range MD simulation\cite{plimp,hendrik} (see also
Ref.\cite{beazley} for a review). To a large extent, this report is based on the work by Plimpton
on short-range MD\cite{plimp}. The first algorithm uses the particle decomposition method wherein
each processor is assigned a fixed number of particles. Each processor calculates the force on the
same particles assigned to it irrespective of its spatial position. The second algorithm uses force
decomposition method in which each processor calculates the sub-block of the force matrix. The
parallelization is implemented using MPI and tested on a system with 4 processors. The purpose of
this report is to present the simplicity of the algorithms in implementing and gain in run time
compared to sequential algorithms on a single processor.

The report is presented as follows : in section II, the computational aspect of the problem is
described. Section III explains the two algorithms in detail, whereas section IV compares the
runtime for the two algorithms with vectorized algorithm.

\section{Basic equation for vortex dynamics and its computational aspect}

A vortex in the mixed phase of a bulk type-II superconductor is a line-like object
which encompasses a quantum of magnetic flux ($\frac{hc}{2e}$ where $h$ is Planck
constant, $c$ is the velocity of light, and $e$ charge of an electron) threaded by
circulating super-current\cite{tinkham}. The interaction between two vortices is
electromagnetic and hence of long range nature. Though the vortex origin requires a
quantum mechanical explanation, most of the physical properties (magnetisation,
transport) of the mixed phase can be described by considering the vortices as
classical interacting objects. In 2D, these vortices can be treated as point massless
particles obeying the following overdamped equation of motion 
\begin{equation} 
\eta \frac{d\vec{r}_{i}}{dt} = \sum_{j\neq i} {\bf F}^{vv}(\vec{r}_{i},\vec{r}_{j}) +
\sum_{j} {\bf F}^{vp} (\vec{r}_{i},\vec{R}_{j}) + {\bf F}^{ext} 
\end{equation} 
where $\vec{r_{i}}$ represents the co-ordinate of the $i$-th particle, and the first
summation runs over all $N-1$ particles. This term represents the repulsive force
between two vortices and is given by the first-order Bessel function $K_{1}(r)$. At
distances $r\gg 1$, the function decays rapidly as $\frac{1}{r^{1/2}}\exp(-r)$. The
second term represents the interaction between the vortex and the quenched disorder.
The disorder co-ordinates $\vec{R}$ is distributed randomly in the simulation region,
and the potential is attractive and short ranged. The last term is the force due to
external condition that can include transport current. The above equation is a
paradigm for driven interacting particles in presence of quenched disorder, and
appears in other fields such charge-density wave (CDW), colloids, tribology.

Computationally, the maximum processor time in a single MD time step goes in
calculating the two-body force represented by the first term. Even with overhead
computational expenses, this could be as high as 95\% (for instance, with 4800
vortices and 2000 point disorder, the first term accounts for 98\% of the processor
time). Therefore, much of the effort has been directed to optimize algorithms for
calculating the two-body force. Since the second term is quenched and short ranged,
it can be calculated efficiently by binning, and summing the forces over the closest
bins. The third term is a one-body term and hence be efficiently vectorized and takes
small fraction of the total computation time. Rest of the report hence deals with the
computational aspect of the first term only.

It is useful to consider the positions of all particles as an $N$-dimensional vector
${\bf r} = (\vec{r}_{1}, \vec{r}_{2}, \vec{r}_{3},........\vec{r}_{N})$ where each of
the $\vec{r}_{i}$ holds the $d$-dimensional co-ordinate of the $i$-th particle
(henceforth, particle is assumed to imply vortex). Thus, for the 2D case $\vec{r}_{i}
= (x_{i},y_{i})$. The force $F_{ij}\equiv F(\vec{r}_{i}, \vec{r}_{j})$ between two
particles at $\vec{r}_{i}$ and $\vec{r}_{j}$ can be thought as an element of a matrix
of size $N\times N$. We denote such a force matrix by {\bf F}. The total force on all
$N$ particles can then be represented as a column vector of length $N$ such that the
$i$-th element is given by $\vec{f}_{i} = \sum_{ij} F_{ij}$. Symbolically, such a
vector {\bf f} can be written as \begin{equation} {\bf f} = {\bf r} \otimes {\bf F}
\cdot {\bf r} \end{equation} The MD simulation thus amounts to calculating ${\bf f}$
at each time step which is then used to update the vector ${\bf r}$.

The computational expense is reflected in the number of elements needed in the summation for
$\vec{f}_{i}$. For long-range forces such as Coulomb force and gravitational force, each particle
interacts with all $(N-1)$ particles, thus requiring $N-1$ elements of $F_{ij}$ in the summation
for $\vec{f}_i$. Therefore, a direct calculation of such forces scales as $N^2$, and becomes
prohibitively expensive even for moderate values of $N$. In the last decade, various approximate
methods have been developed which allows considerable run time gain. They include particle-mesh
algorithms which scale as $f(M)N$ where $M$ is the number of mesh points, hierarchical methods
which goes as $O(N\log N)$\cite{barnes}, and fast multipole methods which scale as $N$\cite{green}.
But most of these algorithms are difficult to implement and requires particular attention to data
structure.

On the other hand, direct methods are specifically suitable for short-range forces in which case
each particle interacts with few particles within a distance $r_c$. Typical examples are MD
simulation employing Lennard-Jones potential and van-der Waals potential. In these simulations, a
significant computational time is spend in searching for the particles within the cut-off distance
$r_c$. There are two techniques usually employed for doing this. The first method constructs list
of all atoms within a distance $r_{c} + \delta$ where $\delta$ is chosen to be small compared to
$r_c$. The list is updated after few timesteps such that the total distance traveled by 
a particle between successive updates is less than $\delta$. The two-body force on a particle is 
found by summing over all particles within the list. The second method, known as link-cell method, 
employs binning of particles at every time step into cells of dimension $r_c$. Since this requires 
$O(N)$ operations, the overhead expenses more than compensates for the time required for 
searching particles within $r_c$ through the list of all $N$ particles. The fastest MD 
algorithms use both these techniques together.

For the interaction under consideration, the inter-particle force decays to a value $\approx 
10^{-8}$ at a distance $r_{c}\approx 15$ (in reduced units). This length scale is considerably 
larger (sometimes nearly half the system size) than usually employed in short-range MD, yet does 
not cover the whole system. Also, since the particle distribution is uniform, this leads to a 
significant fraction of total
number of particles to be within the distance $r_c$ (for example, for 1200 particles in a square
box of length 36, there are on an average 200 particles within a distance $r_c$). The 
construction of neighbour list is therefore of not much advantage and can occupy large memory. 
Also in the situation where particles are driven, the position changes randomly which would 
require frequent updating of the list. If $r_c$ is as large as half the length of the 
simulation box, binning of particles is not expected to give significant time gain (on the 
other hand, this is used for calculating the force due to disorder which is short ranged). The 
MD simulation presented here therefore does not employ both these techniques.

\section{Parallel algorithms}

In the last decade, with the arrival of multi-processor machines, much effort has been spend in
constructing parallel MD algorithms. It is said that the MD computations are inherently parallel in
the sense that force calculation and updates can be done simultaneously on all particles. The aim
of parallelizing MD is to distribute the task of force calculation evenly among all processors. The
various parallel algorithms are based on two basic methods. In the first class of methods, a
pre-determined set of computation is assigned to each processor that remains fixed throughout the
simulation time. One way of doing is to assign a group of particles to each processor, and at any
time the processor computes the force for the particles assigned to it no matter where the
particles are located in the simulation region. This is usually referred as {\bf atom
decomposition} (AD) or {\bf replicated-data} method. Another possible way is to subscribe
sub-blocks of the force matrix computation to each processor that has led to {\bf
force-decomposition} (FD) algorithms. Both these methods are analogous to Lagrangian gridding in
computational fluid dynamics (CFD) where the grid cells (computational elements or processors) move
along with the fluid (particles or atoms in MD). These two algorithms are discussed in detail
below.

The second class of methods uses {\bf spatial decomposition} (SD) wherein the simulation region is 
decomposed into cells and each cell is assigned to a processor. Each processor then computes the 
force on the particles within the cell region assigned to it. This is analogous to Eulerian 
gridding in CFD where the grid remains fixed and the fluid moves through it. In this report, we deal only with 
the implementation of {\it atom decomposition} and {\it force-decomposition} algorithms. Due to 
small number of processors available for computation, the {\it spatial decomposition} algorithm 
cannot be effectively compared with the others.

For the purpose of further discussions, we assume following symbols: the column vectors {\bf r},
{\bf f} and the matrix {\bf F} assigned to a processor with rank $k$ is designated by the subscript
$k$. The rank for the processors are indexed from 0 to $P-1$ where $P$ is the total number of
processors. Thus, ${\bf r}_{k}$ and ${\bf f}_{k}$ are the position vector and total force vector
held in processor $k$, and ${\bf F}_{k}$ is the sub-block of the force matrix assigned to it. The
number of particles (and hence the length of the vector ${\bf r}_{k}$ and ${\bf f}_{k}$) assigned
to the $k$-th processor is represented by $N_{k}$. In all the cases considered here $N_{k} = N/P$.
For simplicity, $N$ is chosen to be an integer multiple of $P$, and is not a constraint on the
algorithm.

\subsection{Atom-Decomposition algorithm}

As mentioned earlier, in this method the computation is carried out by each of the $P$ processors
for $N_{k}(=N/P)$ particles which are assigned to it at the start of the simulation. This amounts
to assigning a sub-block of $N_{k}$ rows of the force matrix ${\bf F}$ to each processors. The
$k$-th processor computes matrix elements of the sub-block ${\bf F}_k$. Let us assume that each
processor has the updated co-ordinates of all particles initially assigned to it. To calculate the
force for the next time step, the particles in $k$-th processor requires positions of all other
particles held by the $P-1$ processors. This implies that at each time step, each of the processor
needs to communicate the co-ordinates of particles held by it to all other processors. This kind of
collective communication can be implemented by calling {\it all-to-all} operation in MPI. An
efficient algorithm to perform the same operation has been given by Fox {\it et al.}\cite{fox} and
is called an {\bf \it expand} operation (see Appendix A for the details).

There are two versions of atom-decomposition method that is discussed in ref.\cite{plimp}. The 
{\bf A1} algorithm is a straightforward implementation of the above mentioned procedure. It 
does not use the skew-symmetric property of the force matrix ${\bf F}$, and hence the two-body 
force $F_{ij}$ is calculated twice (one by the processor holding $\vec{r}_i$ in the sub-vector, 
and the other holding $\vec{r}_j$). The {\bf A1} algorithm is outlined below :

\begin{enumerate}
\item {\bf Expand} vector ${\bf r}_{k}$ and construct ${\bf r}$ in each processor. 
\item {\bf Compute} the sub-block ${\bf F}_{k}$ using ${\bf r}_k$ and ${\bf r}$, and sum the elements into ${\bf f}_k$ in each processor. 
\item {\bf Update} the sub-vector ${\bf r}_k$ in each processor using ${\bf f}_k$.
\end{enumerate}

The communication cost of an algorithm is gauged by the number of messages (in MPI, a pair of
$\textsf{MPI\_SEND}$ and $\textsf{MPI\_RECV}$) and the volume of data exchanged. It would be
fruitful to see both these factors for the above algorithm in each step. The first step expands the
${\bf r}_{k}$ in each processor to construct the full vector ${\bf r}$. This uses $N_k$ vector
length for communication to $P-1$ processors, and therefore scales as $N_{k}P=N$ (actually $N_{k}P
- N_{k}$). The second step computes the force on all the $N_k$ particles in ${\bf r}_k$. This
operation scales as $N\cdot N_{k}=\frac{N^2}{P}$. And the last operation updates the vector ${\bf
r}_{k}$ in each processor which also scales as $N_k$. Thus overall, the communication cost
increases as $N$ and computation cost as $\frac{N^{2}}{P}$. Symbolically, the steps in the above
algorithm can be represented as \begin{eqnarray} {\bf r} & = & E_{P} [{\bf r}_{k}] \nonumber \\
{\bf f}_{k} & = & {\bf r}_{k} \otimes {\bf F} \cdot {\bf r} \end{eqnarray} where ${\bf E}_{P}$ is
an expand operation discussed in Appendix A. Note that ${\bf f}_{k}$ is of length $N_k$.

The second algorithm, denoted by {\bf A2}, uses Newton's law to avoid double computing of the
two-body force. Though this leads to extra communication between processors, this can still
outperform the algorithm {\bf A1} even for moderate values of $N$ when number of processors $P$ is
small. The {\bf A2} algorithm references the pair-wise interaction only once by using a modified
force matrix $G$\cite{plimp}, which is defined as follows : $G_{ij} = F_{ij}$, except that $G_{ij}
= 0$ when $i + j$ is even for $i > j$, and also when $i + j$ is odd for $i < j$. This makes the
matrix $G$ appear as a checkerboard (with the diagonal also set to zero) as shown in Fig.1. The
difference between {\bf A1} and {\bf A2} is in the second step. The force between $i$-th and $j$-th
particle is calculated, and summed into both $i$ and $j$ positions of the resulting force vector
${\bf f}^{'}$. For {\it e.g.}, from Fig.1, at the end of the force calculation, the processor 0 and
1 will have following elements in the vector ${\bf f}^{'}$ 
\begin{minipage}[t]{2.7in}
\begin{center} Processor 0 \end{center} \begin{flushleft}
$F_{1,3}+F_{1,5}+F_{1,7}+F_{1,9}+F_{1,11}-F_{2,1}$ \\
$F_{2,1}+F_{2,4}+F_{2,6}+F_{2,8}+F_{2,10}+F_{2,12}-F_{3,2}$ \\
$F_{3,2}+F_{3,5}+F_{3,7}+F_{3,9}+F_{3,11}-F_{1,3}$ \\ $-F_{2,4}$ \\ $-F_{1,5}-F_{3,5}$ \\
$-F_{2,6}$ \\ $-F_{1,7}-F_{3,7}$ \\ $-F_{2,8}$ \\ $-F_{1,9}-F_{3,9}$ \\ $-F_{2,10}$ \\
$-F_{1,11}-F_{3,11}$ \\ $-F_{2,12}$ \end{flushleft} \end{minipage} \ \ \begin{minipage}[t]{2.7in}
\begin{center} Processor 1 \end{center} \begin{flushright} $-F_{4,1}-F_{6,1}$ \\ $-F_{5,2}$ \\
$-F_{4,3}-F_{6,3}$ \\ $F_{4,1}+F_{4,3}+F_{4,6}+F_{4,8}+F_{4,10}+F_{4,12}-F_{5,4}$ \\
$F_{5,2}+F_{5,4}+F_{5,7}+F_{5,9}+F_{5,11}-F_{6,5}$ \\
$F_{6,1}+F_{6,3}+F_{6,5}+F_{6,8}+F_{6,10}+F_{6,12}-F_{4,6}$ \\ $-F_{5,7}$ \\ $-F_{4,8}-F_{6,8}$ \\
$-F_{5,9}$ \\ $-F_{4,10}-F_{6,10}$ \\ $-F_{5,11}$ \\ $-F_{4,12}-F_{6,12}$ \end{flushright}
\end{minipage} \linebreak\linebreak

This means that the vector obtained as an output from the force calculation in each processor is
again of length $N$ unlike {\bf A1} where it is of size $N_{k}$. The vector ${\bf f}^{'}$ is
summed across all processors so that total force is obtained as the sub-vector ${\bf f}_k$ in each
processor. This operation of summing is called {\bf \it fold} and can be performed optimally by
Fox's algorithm\cite{fox} (see Appendix B). The {\bf A2} algorithm can then be enumerated as
follows :

\begin{enumerate}
\item {\bf Expand} vector ${\bf r}_k$ and construct ${\bf r}$ in each processor.
\item {\bf Compute} the sub-block of $G$ using ${\bf r}_k$ and ${\bf r}$, and obtain ${\bf f}^{'}$.
\item {\bf Fold} vector ${\bf f}^{'}$ across all processors and obtain ${\bf f}_{k}$.
\item {\bf Update} the sub-vector ${\bf r}_k$ in each processor using ${\bf f}_k$.
\end{enumerate}

Note that there is an additional communication (in step 3) of order $N$ as compared to {\bf A1} at
the expense of halving the force calculation. A gain in runtime is possible if the communication
time is a small percentage of overall time which is not the case if $P$ is large. As pointed out by
Plimpton, in that case {\bf A1} algorithm is faster than {\bf A2}. But for the case where the force
calculation is intensive and $P$ is small, {\bf A2} can outperform {\bf A1} which is indeed
observed in our simulation. Symbolically, the {\bf A2} algorithm becomes \begin{eqnarray} {\bf r} &
= & E_{P} [{\bf r}_{k}] \nonumber \\ {\bf f}^{'} & = & {\bf r}_{k} \otimes {\bf G} \cdot {\bf r}
\nonumber \\ {\bf f}_{k} & = & {\bf R}_{P} [{\bf f}^{'}] \end{eqnarray} where ${\bf R}_{P}$ is a
fold operation discussed in Appendix B.

Both {\bf A1} and {\bf A2} algorithms has communication operations which scales as $N$. This for
large values of $N$ can be prohibitive. The advantage with these algorithms is its simplicity,
which allows parallelizing an existing vectorized code (and this was indeed the reason why
parallelization was attempted in our simulation).

\subsection{Force-Decomposition algorithm}

The force-decomposition (FD) algorithm, which was first implemented by Hendrickson and Plimpton for
short-range forces, was motivated by block-decomposition of matrix for solving linear algebra
problems. An important aspect of this algorithm is that the communication between processors scales
as $N/\sqrt{P}$ rather than $N$ for the atom-decomposition algorithm. For the FD algorithm the
processors are assumed to be arranged in cartesian topology (see ref.\cite{mpi}). Also, we use the
following notations : the cartesian co-ordinates of the processor is denoted by the subscripts
$(j,k)$ ($j$ is a row index, and $k$ the column index) and runs from $0$ to $\sqrt{P}-1$. Thus, 
${\bf r}_{j,k}$ and ${\bf f}_{j,k}$ are the position and total force vectors held by the processor 
with co-ordinates $(j,k)$. We also use notation in which row (column) index in the subscript is 
absent which implies that the same vector is held by all processors in the given column (row) 
index. As an example, ${\bf r}_{j,}$ is held by all processors in the $j$-th row, and ${\bf 
r}_{,k}$ by all processors in the $k$-th column. This typically arises when a vector is expanded 
along a row or column of processors (see appendix A). Also, for simplicity we assume that $P$ is an 
even power of $2$, and $N$ is an multiple of $P$.

The block decomposition of the force matrix is performed not on ${\bf F}$ but on a new matrix ${\bf
F}^{'}$ obtained by using vector ${\bf r}$ and permuted vector ${\bf r}^{'}$. The permuted vector
${\bf r}^{'}$ is formed by lining up ${\bf r}_{j,k}$ held by each processors column wise {\it
i.e.}, ${\bf r}^{'} = ({\bf r}_{0,0}, {\bf r}_{1,0},......{\bf r}_{\sqrt{P}-1,0}, {\bf r}_{0,1},
{\bf r}_{1,1},....,{\bf r}_{\sqrt{P}-1,1},......)$. Note that the vector ${\bf r}$ is distributed
as ${\bf r} = ({\bf r}_{0,0}, {\bf r}_{0,1},......{\bf r}_{0,\sqrt{P}-1}, {\bf r}_{1,0}, {\bf
r}_{1,1},......{\bf r}_{1,\sqrt{P}-1},......)$. Both these vectors are shown in Fig.2. The elements
of the force matrix $F^{'}_{ij}$ is the force between particles at $\vec{r}_{i}$ and
$\vec{r}^{'}_{j}$.

To calculate the elements of the sub-block ${\bf F}^{'}_{j,k}$, two sub-vectors of length
$N/\sqrt{P}$ each from ${\bf r}$ and ${\bf r}^{'}$ are required. For a processor with cartesian
co-ordinates $(j,k)$, these vectors are in fact obtained by expand operation across $j$-th row and
$k$-th column. These vectors are represented as ${\bf r}_{j,}$ and ${\bf r}^{'}_{,k}$. As an
example, from Fig.2, the vectors ${\bf r}_{j,}$ and ${\bf r}^{'}_{,k}$ are given below for all 
the 4 processors : 
\begin{flushleft} 
Processor 0 : ${\bf r}_{0,}$ = \{1,2,3,4,5,6\} and ${\bf r}_{,0}$ = \{1,2,3,7,8,9\} \\ 
Processor 1 : ${\bf r}_{0,}$ = \{1,2,3,4,5,6\} and ${\bf r}_{,1}$ = \{4,5,6,10,11,12\} \\ 
Processor 2 : ${\bf r}_{1,}$ = \{7,8,9,10,11,12\} and ${\bf r}_{,0}$ = \{1,2,3,7,8,9\} \\ 
Processor 3 : ${\bf r}_{1,}$ = \{7,8,9,10,11,12\} and ${\bf r}_{,1}$ = \{4,5,6,10,11,12\} 
\end{flushleft}

The force is then calculated between the elements of ${\bf r}_{j,}$ and ${\bf r}^{'}_{,k}$, and
summed in a sub-vector ${\bf f}^{'}_{j,k}$ which is of length $N/\sqrt{P}$. The total force
sub-vector ${\bf f}_{j,k}$ is obtained by a fold operation on ${\bf f}^{'}_{j,k}$ across the $j$-th
row of processors. Note that the expand and fold operations are along a row or column requiring
only $\sqrt{P}$ processors, unlike AD algorithm where all $P$ processors are involved. This leads
to an overall communication cost which scales as $N/\sqrt{P}$ only unlike $N$ in AD algorithms. The
algorithm is summarised in the following steps :

\begin{enumerate}
\item {\bf Expand} vector ${\bf r}_{j,k}$ along $j$-th row, and construct ${\bf r}_{j,}$ in each processor.
\item {\bf Expand} vector ${\bf r}_{j,k}$ along $k$-th column, and construct ${\bf r}^{'}_{,k}$ in each processor.
\item {\bf Compute} the sub-block of ${\bf F}^{'}_{j,k}$ using ${\bf r}_{j,}$ and ${\bf r}^{'}_{,k}$, and obtain ${\bf f}^{'}_{j,k}$.
\item {\bf Fold} vector ${\bf f}^{'}_{j,k}$ across $j$-th row of processors and obtain ${\bf f}_{j,k}$.
\item {\bf Update} the sub-vector ${\bf r}_{j,k}$ in each processor using ${\bf f}_{j,k}$.
\end{enumerate}

Symbolically, the first four steps of the above algorithm can be represented as 
\begin{eqnarray}
{\bf r}_{j,} & = &{\bf E}_{P_{,k}}[{\bf r}_{j,k}] \nonumber \\ 
{\bf r}^{'}_{,k} & = & {\bf E}_{P_{j,}}[{\bf r}_{j,k}] \nonumber \\ 
{\bf f}^{'}_{j,k} & = & {\bf r}_{j,} \otimes {\bf F}^{'}
\cdot {\bf r}_{,k}^{'} \nonumber \\ 
{\bf f}_{j,k} & = & {\bf R}_{P_{j,}} [{\bf f}^{'}_{j,k}]
\end{eqnarray} 
Note that there are 3 communication processes, 2 expand, and 1 fold operation involved in this 
algorithm.

This algorithm does not make use of Newton's law and is designated as {\bf F1} algorithm. Using
Newton's law leads to halving of the computation at the expense of increased inter-processor
communication. Instead of using ${\bf F}$ for obtaining a permuted matrix, one can use the
checkerboard matrix ${\bf G}$ described for {\bf A2} algorithm. Such a matrix is shown in Fig. 3
for $P=4$ and $N=12$. Using {\bf G} modifies the {\bf F1} algorithm in step 3 and 4. In step 3, the
elements of ${\bf G}^{'}_{j,k}$ are computed and accumulated in $i^{th}$ position of ${\bf
f}^{'}_{j,k}$ and $j^{th}$ position of ${\bf f}^{''}_{j,k}$. For example, these vectors are 
given explicitly below for the two processors shown in Fig.3 :

\begin{center}
Processor 0
\end{center}
\begin{minipage}[t]{2.7in}
${\bf f}^{'}_{0,0}$  \\
$F_{1,3}+F_{1,7}+F_{1,9}$ \\
$F_{2,1}+F_{2,8}$ \\
$F_{3,2}+F_{3,7}+F_{3,9}$ \\
$F_{4,1}+F_{4,3}+F_{4,8}$ \\
$F_{5,2}+F_{5,7}+F_{5,9}$ \\
$F_{6,1}+F_{6,3}+F_{6,8}$ \\
\end{minipage} \  \
\begin{minipage}[t]{2.7in}
${\bf f}^{''}_{0,0}$ \\
$F_{2,1}+F_{4,1}+F_{6,1}$ \\
$F_{3,2}+F_{5,2}$ \\
$F_{1,3}+F_{4,3}+F_{6,3}$ \\
$F_{1,7}+F_{3,7}+F_{5,7}$ \\
$F_{2,8}+F_{4,8}+F_{6,8}$ \\
$F_{1,9}+F_{3,9}+F_{5,9}$ \\
\end{minipage} 
\linebreak
\begin{center}
Processor 1
\end{center}
\begin{minipage}[t]{2.7in}
${\bf f}^{'}_{0,1}$ \\
$F_{1,5}+F_{1,11}$ \\
$F_{2,4}+F_{2,6}+F_{2,10}+F_{2,12}$ \\
$F_{3,5}+F_{3,11}$ \\
$F_{4,6}+F_{4,10}+F_{4,12}$ \\
$F_{5,4}+F_{5,11}$ \\
$F_{6,5}+F_{6,10}+F_{6,12}$ \\
\end{minipage} \  \
\begin{minipage}[t]{2.7in}
${\bf f}^{''}_{0,1}$  \\
$F_{2,4}+F_{5,4}$ \\
$F_{1,5}+F_{3,5}+F_{6,5}$ \\
$F_{2,6}+F_{4,6}$ \\
$F_{2,10}+F_{4,10}+F_{6,10}$ \\
$F_{1,11}+F_{3,11}+F_{5,11}$ \\
$F_{2,12}+F_{4,12}+F_{6,12}$ \\
\end{minipage} 

The ${\bf f}^{'}_{j,k}$ is folded across the row of processors to obtain ${\bf f}^{'}_{j,}$, and
${\bf f}^{''}_{j,k}$ is folded across the column of processors to obtain ${\bf f}^{''}_{,k}$. The
total force ${\bf f}_{j,k}$ is obtained by subtracting ${\bf f}^{'}_{j,}$ from ${\bf f}^{''}_{,k}$
element by element.  In the above example, folding the vector ${\bf f}^{'}_{0,0}$ and ${\bf
f}^{'}_{0,0}$ along the row gives for particle 1 
\begin{center}
$\underbrace{F_{1,3}+F_{1,7}+F_{1,9}}_{{\bf f}^{'}_{0,0}}+\underbrace{F_{1,5}+F_{1,11}}_{{\bf
f}^{'}_{0,1}} 
$ \end{center} 
whereas same operation on ${\bf f}^{''}_{0,0}$ and ${\bf f}^{''}_{1,0}$ along the column gives 
\begin{center} 
$\underbrace{F_{2,1}+F_{4,1}+F_{6,1}}_{{\bf 
f}^{''}_{0,0}}+\underbrace{F_{8,1}+F_{10,1}+F_{12,1}}_{{\bf f}^{''}_{1,0}} $ 
\end{center} 
The total force on particle 1 is obtained by subtracting the second sum from the first. This also 
brings out an important feature that all elements of the force matrix for an $i$-th particle is 
calculated by those processors which share the row and column with the processor to which $i$-th 
particle is assigned. The {\bf F2} algorithm can be enumerated as follows :

\begin{enumerate}
\item {\bf Expand} vector ${\bf r}_{j,k}$ along $j$-th row, and construct ${\bf r}_{j,}$ in each processor.
\item {\bf Expand} vector ${\bf r}_{j,k}$ along $k$-th column, and construct ${\bf r}^{'}_{,k}$ in each processor.
\item {\bf Compute} the sub-block ${\bf G}^{'}_{j,k}$ and accumulate them in ${\bf f}^{'}_{j,k}$ and ${\bf f}^{''}_{j,k}$.
\item {\bf Fold} vector ${\bf f}^{'}_{j,k}$ across $j$-th row of processors and and obtain ${\bf f}^{'}_{j,}$.
\item {\bf Fold} vector ${\bf f}^{''}_{j,k}$ across $k$-th column of processors and obtain ${\bf f}^{''}_{,k}$.
\item {\bf Subtract} ${\bf f}^{''}_{,k}$ from ${\bf f}^{'}_{j,}$, and obtain ${\bf f}_{j,k}$
\item {\bf Update} the sub-vector ${\bf r}_{j,k}$ in each processor using ${\bf f}_{j,k}$.
\end{enumerate}

Symbolically, the {\bf F2} algorithm has the following structure
\begin{eqnarray}
{\bf r}_{j,} & = &{\bf E}_{P_{,k}} [{\bf r}_{j,k}] \nonumber \\
{\bf r}^{'}_{,k} & = & {\bf E}_{P_{j,}} [{\bf r}_{j,k}] \nonumber \\
{\bf f}^{'}_{j,k} & = & {\bf r}_{j,} \otimes {\bf G}^{'} \cdot {\bf r}_{,k}^{'} \nonumber \\
{\bf f}^{'}_{j,} & = & {\bf R}_{P_{j,}} [{\bf f}^{'}_{j,k}] \nonumber \\
{\bf f}^{''}_{,k} & = & {\bf R}_{P_{,k}} [{\bf f}^{''}_{j,k}] \nonumber \\
{\bf f}_{j,k} & = & {\bf f}^{'}_{j,} - {\bf f}^{''}_{,k} \nonumber \\
\end{eqnarray}
The increased communication and computation is evident in step 5 and 6 at the expense of reduced force computation.

\section{Results}

We have implemented all the four algorithms {\bf A1, A2, F1} and {\bf F2} on a 4 processor (each of
180MHz) SGI system. The first order differential equation (1) is solved using a fourth order
predictor-corrector scheme. This requires evaluating the right hand side of Eq.(1) twice at each
time step. All the calculation is done in double precision. Since the force between two particles
is Bessel function of order 1 ($K_{1}(r)$), evaluating the function at each time step is expensive.
We follow the usual practice of tabulating the function at regular interval over the full range.
The force at any distance is then obtained by interpolating the values in the table.  The program
is written in Fortran 90 whereas the inter-processor communication is handled using MPI. For the AD
algorithms, the default communicator world is used which does not specify any topology for the
processors. For FD algorithms, the cartesian topology is imposed leading to a grid of $2\times 2$
processors. The number of particles tested ranged from 120 to 4800. The idea here is not to
simulate large particles but to show the possibility of parallel computation and its advantage even
on a small number of processors.

Fig. 4(a)  shows the CPU time required for a single MD time step with increasing density of
particles for different algorithms using all 4 processors. The time for {\bf A1} algorithm and {\bf
F1} is nearly equal. Note that both these algorithms computes the inter-particle force twice. The
number of communications for {\bf A1} is less than that for {\bf F1} at the expense of increased
force computation. On the other hand, for {\bf A2} and {\bf F2} algorithms, the number of
communications are same, and {\bf A2} outperforms {\bf F2} due to less computation required. Note
that with increasing number of processors, {\bf F2} is expected to perform better \cite{hendrik} as
communication cost for this algorithm goes as $O(N/\sqrt{P})$ rather than $O(N)$ for {\bf A2}.

Though, the total number of processors available is too small to obtain the trends with varying
number of processors, it would still be instructive to plot the CPU time as a function of $P$
for a system of 4800 particles and is shown in Fig.4(b). The dotted line represents the ideal 
speed up, which is obtained by assuming that the time required for a single processor is 
equally divided among all available processors.

\section {Conclusions}

This report shows that considerable gain in run time can be achieved on implementing parallel MD
simulation even for a moderate number of processors. Though the algorithms that have been
implemented here were used by others for short-range MD, present work shows that it can be used
even for forces that are inter-mediate range with advantage over vectorized algorithms. As stressed
in this report, the main advantage is its simplicity in implementing and requiring few 
modifications for an already vectorized simulation code for a sequential machine.

For this work, the emphasis is simply on the parallelization of the basic MD simulation. In the
implementation reported here, a significant fraction of the computing time goes in searching for
particles within the cut-off distance for the force. It is to be seen as to how the conventional
methods of MD, which uses nearest neighbour tables and link-cell method would enhance the
performance. As stressed before, the problem at hand is that of driven particles in presence of
quenched random disorder. At depinning velocities when a fraction of the particles move, the
neighbourhood of a particle changes rapidly with time. This would call for frequent updating of the
nearest neighbour table. Also, maintaining such a table for long range force may affect the overall
performance which need to be verified. Other approximate methods for long-range force
\cite{barnes,green} need to be explored though these algorithms are difficult to 
implement on distributed memory machines.

\section*{Appendix A}

The {\it expand} operation on vectors ${\bf x}_{k}=\{x_{i} \}_{i=1}^{N_{k}}$ (the rank of the
processor is denoted by subscript $k$) held by $P$ processors results in a vector ${\bf y} = \{
y_{i} \}_{i=1}^{N}$ in all processors such that ${\bf y} = ({\bf x}_{1}, {\bf x}_{2}, {\bf
x}_{3},..... {\bf x}_{P})$. The length of the vector ${\bf y}$ in all the $P$ processors is
$N=N_{k}P$. Thus, vector ${\bf y}$ is constructed by arranging vector ${\bf x}$ in ascending order
of the rank of the processors. Symbolically, this operation can be represented as 
\begin{equation}
{\bf E}_{P} [{\bf x}_{k}] = {\bf y} 
\end{equation} 
where ${\bf x} = \{ x_{i}\}_{i=1}^{N_{k}}$ and ${\bf y} = \{y_{i}\}_{i=1}^{N_{k}P}$.  Thus after an 
{\it expand} operation, the output buffer in each processor has a vector of length which is $P$ 
times the length of the vector held by each processor.

For a 2D grid of processors, if $(j,k)$ is the cartesian co-ordinates, an expand operation across
$j$-th row can be symbolically represented as 
\begin{equation} 
{\bf E}_{P_{,k}} [{\bf x}_{j,k}] = {\bf y}_{j,} 
\end{equation} 
and that along the $k$-th column as 
\begin{equation} 
{\bf E}_{P_{j,}} [{\bf x}_{j,k}] = {\bf y}_{,k} 
\end{equation} 
The expand operation along the row (column) returns a vector of length $N_{k}$ time  the number 
of processors in a column (row).

Fox {\it et al}.\cite{fox} has prescribed an elegant and simple method to perform this operation
(also see ref.\cite{plimp}). In the first step, each processor allocates memory for the full vector
${\bf y}$, and the vector ${\bf x}_{k}$ is mapped to its position in ${\bf y}$. Thus, at the
beginning of the {\it expand}, $k$-th processor has ${\bf y} = (0, 0,....{\bf x}^{k},...0)$. In the
first communication step, each processor partners with the neighbouring processor, and exchange
non-zero sub-pieces. At the end of this step, ${\bf y} = (0, 0,....{\bf x}^{k}, {\bf x}^{k+1}...0)$
for the $k$-th and $(k+1)$-th processors. Every processor thus obtains a contiguous vector of
length $2N_{k}$. In the next step, every processor partners with a processor that is two positions
away, and exchanges the new piece (of length $2N_{k}$). This leads to ${\bf y} = (0, 0,....{\bf
x}^{k}, {\bf x}^{k+1}, {\bf x}^{k+2}, {\bf x}^{k+3}....0)$ in each of $k, k+1, k+2, k+3$
processors. This steps leads to each processor acquiring $4N_{k}$ contiguous vector length. This
procedure is repeated till each processor communicates with a processor that is $P/2$ position
away, and at the end of which the entire vector {\bf y} is acquired by all the $P$ processors. Fig.
A1 shows the procedure for $P=4$ and $N_{k}=3$.

For the {\it expand} operation given above, there are $\log_{2} P$ messages and $N_{k}P - N_{k}$
length of vector exchanged. This is an optimal value. Allocation of $N_{k}P$ memory in each of the
processing element can become prohibitive when $N_{k}$ and $P$ are large, though for most purpose
this is never a real concern. Note that the number of processors involved in {\it expand} operation
can be a sub-set of total number of available processors (see FD algorithms).

\section*{Appendix B}

The {\it fold} operation between $P$ processors is the inverse of {\it expand}. If each processor
holds a vector ${\bf y} = \{y_{i} \}_{i=1}^{N}$ where $N=N_{k}P$, the {\it fold} operation between
$P$ processors leads to $k$-th processor acquiring a vector ${\bf x} = \{ x_{i} \}_{i=1}^{N_k}$.
The vector ${\bf x}$ is constructed by summing up vector ${\bf y}$ across all $P$ processors and
allocating the $k$-th segment of it. Thus, if ${\bf y}^{j} = ({\bf y}^{j}_{1}, {\bf y}^{j}_{2},
{\bf y}^{j}_{3},..... {\bf y}^{j}_{P})$ where ${\bf y}^{j}_{k}$ is the $k$-th segment of ${\bf y}$
(of length $N_{k}$) in $j$-th processor, after a {\it fold} operation the $k$-th processor receives
${\bf x}_{k} = \sum_{j=1}^{P} {\bf y}_{k}^{j}$. Symbolically, we represent this operation as
\begin{equation} 
{\bf R}_{P} [{\bf y}] = {\bf x}_{k} 
\end{equation} where ${\bf y} = \{y_{i}\}_{i=1}^{N}$ and ${\bf x} = \{x_{i}\}_{i=1}^{N/P}$. Thus 
after a {\it fold} operation, the output buffer has a vector of length $N/P$ where $N$ is the 
vector length used for {\it folding}.

The {\it expand} algorithm by Fox {\it et al}. can be inverted to obtain a simple and optimal
method to perform {\it fold} operation (also see ref.\cite{plimp}). In the first step, each
processor pairs up with processor that is $P/2$ position away, and exchanges pieces of vector
length $N/2$. The pieces that are received are the one that the processor must acquire as a final
sum. Thus, processor 0 pairs up with $P/2$ and sends vector segment ${\bf y} =
\{y_{i}\}_{i=\frac{N}{2}+1}^{N}$ and receives ${\bf y} = \{y_{i}\}_{i=1}^{N/2}$, whereas the
$P/2$-th processor sends vector segment ${\bf y} = \{y_{i}\}_{i=1}^{N/2}$ and receives ${\bf y} =
\{y_{i}\}_{i=\frac{N}{2}+1}^{N}$. In the next step, the processor pairs up with the processor at
$P/4$ and exchanges $N/4$ length of the new vector ${\bf y}$, again noting that each processor in
the pair receives that part which it requires in the summation. This procedure is done recursively,
and in each step the vector length exchanged is halved. Fig. A2 shows the entire procedure for
$P=4$ and $N=12$. This algorithm is again an optimal one requiring $\log_{2}P$ steps and requiring
exchange of $N - N/P$ length of vector.

\pagebreak
\section*{Figure Captions}
\newcounter{bean}
\begin{list}%
{Fig.\arabic{bean}}{\usecounter{bean}}
\item The checkerboard matrix used for computing in {\bf A2} algorithm. 

\item (a) The arrangement of processors and the vectors {\bf r} and ${\bf r}^{'}$ for {\bf F1} and
{\bf F2} algorithms. Note that the vector ${\bf r}^{'}$ is obtained by expanding first along column
and then across the row, whereas the original vector {\bf r} is generated by first expanding across
a row and then along a column. The block-decomposition of the force matrix is shown in (b).

\item The block-decomposition of the force matrix for {\bf F2} algorithm. Note the vector on the
top ${\bf r}^{'}$ and compare it with that in Fig.1. The permuted vector leads to change in the
checkerboard arrangement of the force matrix.

\item (a) The CPU time taken for a single MD step by different algorithms for a 4 processor system.

(b) The CPU time taken for a single MD step by AD and FD algorithms with increasing number of
processors in a simulation of 4800 particles . The dotted line is the ideal speed up and is the
time for a single processor divided by the number of processors.

\end{list}

\newcounter{bean1} 
\begin{list}%
{Fig. A\arabic{bean1}}{\usecounter{bean1}} 
\item The {\it expand} operation across 4 processors ranked 0 to 3 for a vector ${\bf x}_{k}$ 
with 3 elements. In the first step, each processor partners with nearest processor and exchanges 
sub-pieces. Thus 0 gives elements \{1,2,3\}, and receives \{4,5,6\} elements of 1. After this, both 
0 and 1 has contiguous array of elements \{1,2,3,4,5,6\}. Similarly 2 and 3 has \{7,8,9,10,11,12\}. 
In the second step, the processor pairs with those situated 2 positions away and repeats the 
exchange of non-zero pieces thus generating the full vector. There are only $\log_{2}P$ steps.

\item The {\it fold} operation is an inverse of {\it expand}. This is shown here for $P=4$ and
${\bf x}_{k}$ with 3 elements (total vector length $N=12$). In the first step, each processor pairs
with the processor that is $P/2$ position away and sends $N/2$ sub-piece in which it is {\bf not} a
member. Thus, processor 0 gets the first half of 2 and sends it the second half. The pieces are
added up and stored in the same vector (the stacking of the pieces sidewise in the figure implies
addition). In the second step, the processor partners with those at position $P/4$ away, and
receives the $N/4$ piece in which it is a member. Thus 0 processor receives the \{1,2,3\} of
processor 1, and sends it the elements \{4,5,6\}. Note that the elements \{1,2,3\} is the sum of
elements held by processor 2 and 4. The {\it fold} operation also requires only $\log_{2}P$ steps.

\end{list}


\begin{references}
\bibitem{plimp} S. Plimpton, {\it J. Comp. Phys.}, {\bf 117}, 1, 1995.
\bibitem{hendrik} B. A. Hendrickson and S. Plimpton, {\it J. Par. and Dist. Comp.}, {\bf 27}, 15, 1995.
\bibitem{beazley} D. M. Beazley, P. S. Lomdahl, N. Gr\o nbech-Jensen, R. Giles, and P. Tamayo, in {\it Annual Reviews of Computational Physics III}, Ed. D. Stauffer, World Scientific, Singapore (1995).
\bibitem{tinkham} M. Tinkham, {\it Introduction to Superconductivity}, Krieger Publishing Company, Florida, 1980.
\bibitem{barnes} J. E. Barnes and P. Hut, {\it Nature}, {\bf 324}, 446, 1986.
\bibitem{green} L. Greengard and V. Rokhlin, {\it J. Comp. Phys.}, {\bf 73}, 325, 1995.
\bibitem{fox} G.C. Fox, M. A. Johnson, G. A. Jyzenga, S. W. Otto, J. K. Salmon, and D. A. Walker, {\it Solving Problems on Concurrent Processors: Volume 1}, Prentice Hall, Englewood Cliffs, NJ, 1988.
\bibitem{mpi} {\it MPI : A Message-Passing Interface Standard}, Ver.1.1, MPI Forum, 1995.
\end{references}
\end{document}